\def \VersionAuthor {}
\ifdefined\VersionAuthor
	\newcommand{\AuthorVersion}[1]{#1}
	\newcommand{\SpringerVersion}[1]{}
\else
	\newcommand{\AuthorVersion}[1]{}
	\newcommand{\SpringerVersion}[1]{#1}
\fi

\documentclass[a4paper,10pt]{llncs}
\let\oldparagraph=\paragraph
\renewcommand\paragraph[1]{\oldparagraph{#1.}}

\usepackage[utf8]{inputenc}
\usepackage[english]{babel}

\usepackage[ruled,vlined,linesnumbered]{algorithm2e}
	\SetKwInOut{Input}{input}
	\SetKwInOut{Output}{output}

\usepackage{subcaption}
	\usepackage[compatibility=false,labelfont=bf]{caption}
\usepackage{paralist} % inline lists

\usepackage{amsmath} % for \text et al., for ``cases'' environment…
\usepackage{amssymb} % for \mathbb et al.

\ifdefined\VersionWithComments
	\usepackage{draftwatermark}
	\SetWatermarkText{draft}
	\SetWatermarkScale{15}
	\SetWatermarkColor[gray]{0.9}
\fi
\usepackage[svgnames,table]{xcolor}
\definecolor{darkblue}{rgb}{0.0,0.0,0.6}
\definecolor{darkgreen}{rgb}{0, 0.5, 0}
\definecolor{darkpurple}{rgb}{0.7, 0, 0.7}
\definecolor{darkblue}{rgb}{0, 0, 0.7}

\usepackage[
		colorlinks=true,
	\ifdefined \VersionWithComments
		pagebackref=true,
	\fi
		citecolor=darkgreen,
		linkcolor=darkblue,
		urlcolor=darkpurple,
	]{hyperref}

\usepackage[capitalise,english,nameinlink]{cleveref} % load after algorithm2e and hyperref
\crefname{line}{\text{line}}{\text{lines}} % to remove the capital

\ifdefined \VersionLong
	\newcommand{\LongVersion}[1]{\ifdefined\VersionWithComments{\color{red!40!black}#1}\else#1\fi}
	
\else
	\newcommand{\LongVersion}[1]{\ifdefined\VersionWithComments{\color{black!40}#1}\fi}
	
\fi

\usepackage{tikz}
\usetikzlibrary{arrows,automata,positioning}
\tikzstyle{every node}=[initial text=]
\tikzstyle{location}=[rectangle, rounded corners, minimum size=12pt, draw=black, fill=blue!10, inner sep=2pt]
\tikzstyle{invariant}=[draw=black, dotted, inner sep=1pt] % xshift=1em, 
\tikzstyle{final}=[double]
\tikzstyle{accepting}=[final]

\newcommand{\styleclock}[1]{\ensuremath{\textcolor{colorclock}{#1}}}

\newcommand{\styleparam}[1]{\ensuremath{\textcolor{colorparam}{#1}}}

\newcommand{\styleBen}[1]{\texttt{#1}}

\newcommand{\stylealgo}[1]{\ensuremath{\textsf{#1}}}
\newcommand{\EFsynth}{\stylealgo{EFsynth}}

\definecolor{coloract}{rgb}{0.50, 0.70, 0.30}
\definecolor{colorclock}{rgb}{0.4, 0.4, 1}
\definecolor{colorconst}{rgb}{0.50, 0.20, 0.00}
\definecolor{colordisc}{rgb}{1, 0, 1}
\definecolor{colorloc}{rgb}{0.4, 0.4, 0.65}
\definecolor{colorparam}{rgb}{1, 0.6, 0.0}

\newcommand{\cellSubHeader}[1]{\cellcolor{blue!10}\textbf{#1}}
\newcommand{\rowHeader}{\rowcolor{blue!20}\bfseries}
\newcommand{\cellYes}{\cellcolor{green!20}\textbf{$\surd$}}
\newcommand{\cellNo}{\cellcolor{red!20}\textbf{$\times$}}

\newcommand{\A}{\ensuremath{\mathcal{A}}}

\newcommand{\Clock}{\mathbb{X}} % set of clocks
\newcommand{\clock}{x} % clock
\newcommand{\grandn}{{\mathbb N}}

\newcommand{\loc}{l} % location

\newcommand{\Param}{\mathbb{P}} % set of parameters (P)
\newcommand{\param}{p} % parameter (p)
\newcommand{\Variable}{\mathbb{V}} % set of discrete

\ifdefined \VersionWithComments
	\usepackage{marginnote}
	\newcommand{\marginX}{\marginnote{\huge{\quad\quad\textbf{!}\quad\quad}}}
	\newcommand{\ea}[1]{\mbox{}{\color{blue}\marginX{}\textbf{[\'Etienne}: #1]}}
	\newcommand{\instructions}[1]{{\color{red}\marginX{}\textbf{[Instructions: ``#1'']}}}
	\newcommand{\reviewer}[2]{\mbox{}{\color{red}\marginX{}\textbf{[Reviewer #1}: ``#2'']}}
	\newcommand{\todo}[1]{\mbox{}{\color{red}{\marginX{}\textbf{TODO}\ifx#1\\\else:\ \fi #1}}} % here, ``\\'' stands for ``empty''
\else
	\newcommand{\instructions}[1]{}
	\newcommand{\ea}[1]{}
	\newcommand{\reviewer}[2]{}
	\newcommand{\todo}[1]{}
\fi

\newcommand{\hytech}{\textsc{HyTech}}
\newcommand{\imitator}{\textsf{IMITATOR}}
\newcommand{\phaver}{\texttt{PHAVer}}
\newcommand{\psyhcos}{PSyHCoS}
\newcommand{\romeo}{\textsc{Romeo}}
\newcommand{\spaceex}{\texttt{SpaceEx}}
\newcommand{\symrob}{\texttt{Symrob}}
\newcommand{\uppaal}{\textsc{Uppaal}}

\ifdefined \VersionWithComments
 	\definecolor{colorok}{RGB}{80,80,150}
\else
	\definecolor{colorok}{RGB}{0,0,0}
\fi

\newcommand{\eg}{\textcolor{colorok}{e.\,g.,}\xspace}
\newcommand{\etal}{\textcolor{colorok}{\emph{et al.}}\xspace}
\newcommand{\ie}{\textcolor{colorok}{i.\,e.,}\xspace}

\newcommand{\wrt}{\textcolor{colorok}{w.r.t.}\xspace}

\title{A benchmark library for parametric timed model checking\thanks{%
	\LongVersion{This is the author %(and slightly extended)
		version of the manuscript of the same name published in the proceedings of the Sixth International Workshop on Formal Techniques for Safety-Critical Systems (FTSCS 2018).
	The final version is available at \url{dx.doi.org/10.1007/XXXXXXXXX}.
	}%
	This work is partially supported by the ANR national research program PACS (ANR-14-CE28-0002)
		and
	by ERATO HASUO Metamathematics for Systems Design Project (No.\ JPMJER1603), JST.
}}
\author{\'Etienne Andr\'e\inst{1,2,3}\orcidID{0000-0001-8473-9555}}
\institute{Université Paris 13, LIPN, CNRS, UMR 7030, F-93430, Villetaneuse, France
\and JFLI, CNRS, Tokyo, Japan
\and National Institute of Informatics, Japan
\SpringerVersion{\\\email{eandre93430@lipn13.fr}}
}

\begin{document}

% For all page numbers, except p.1
\AuthorVersion{
	\pagestyle{plain}
}

\maketitle
% HACK: because of LNCS style!
\setcounter{footnote}{0}

% For page numbers on p.1
\AuthorVersion{
	\thispagestyle{plain}
}

\ifdefined \VersionWithComments
	\textcolor{red}{\textbf{This is the version with comments. To disable comments, comment out line~3 in the \LaTeX{} source.}}
\fi

\begin{abstract}
	Verification of real-time systems involving hard timing constraints and concurrency is of utmost importance.
	Parametric timed model checking allows for formal verification in the presence of unknown timing constants or uncertainty (\eg{} imprecision for periods).
	With the recent development of several techniques and tools to improve the efficiency of parametric timed model checking, there is a growing need for proper benchmarks to test and compare fairly these tools.
	We present here a benchmark library for parametric timed model checking made of benchmarks accumulated over the years.
	Our benchmarks include academic benchmarks, industrial case studies and examples unsolvable using existing techniques.
\end{abstract}

\instructions{
We solicit submissions reporting on:
\begin{itemize}
\item [A] original research contributions (16 pages max, LNCS format);
\item [B] applications and experiences (16 pages max, LNCS format);
\item [C] surveys, comparisons, and state-of-the-art reports (16 pages max, LNCS);
\item [D] tool papers (5 pages max, LNCS format);
\item [E] position papers and work in progress (5 pages max, LNCS format)
\end{itemize}
}

\todo{uppaal: stopwatches / insister sur les études de cas impossibles à résoudre qui orientent la recheche}

%%%%%%%%%%%%%%%%%%%%%%%%%%%%%%%%%%%%%%%%%%%%%%%%%%%%%%%%%%%%
%%%%%%%%%%%%%%%%%%%%%%%%%%%%%%%%%%%%%%%%%%%%%%%%%%%%%%%%%%%%
\section{Introduction}\label{section:introduction}
%%%%%%%%%%%%%%%%%%%%%%%%%%%%%%%%%%%%%%%%%%%%%%%%%%%%%%%%%%%%
%%%%%%%%%%%%%%%%%%%%%%%%%%%%%%%%%%%%%%%%%%%%%%%%%%%%%%%%%%%%

Verification of real-time systems involving hard timing constraints and concurrency is of utmost importance, and is now recognized in standards such as the DO-178C, that allows formal methods without addressing specific process requirements.
Model checking is a popular model-based technique that formally verifies whether a model satisfies a property.
Parametric timed model checking significantly enhances model checking by allowing its application earlier in the design phase, when timing constants may not be known yet.
In addition, it is possible to verify systems in the presence of uncertainty, \eg{} when some periods are known with some limited precision.
This is the case of Thales' FMTV\footnote{%
	``Formal Methods for Timing Verification Challenge'', in the WATERS workshop: \url{http://waters2015.inria.fr/}
} challenge 2014 where the system was characterized with uncertain but \emph{constant} periods, that rules out the use of non-parametric timed model checking.

Popular formalism for parametric timed model checking include parametric timed automata (PTAs)~\cite{AHV93} and parametric time Petri nets~\cite{TLR09}.

Several tools support parameters, such as
	\hytech{}~\cite{HHW95} (parametric hybrid automata),
	\romeo{}~\cite{LRST09} (parametric time Petri nets),
	\imitator{}~\cite{AFKS12} (parametric timed automata),
	\psyhcos{}~\cite{ALSDL13} (parametric stateful timed CSP),
	or
	\symrob{} (robustness for timed automata)~\cite{Sankur15}.
In addition, several tools support the larger class of \emph{hybrid automata}, such as \phaver{}~\cite{Frehse2008} or \spaceex{}~\cite{FLDCRLRGDM11} and, while not explicitly supporting parameters, can encode them.\footnote{%
	In a hybrid automaton, a parameter is a variable that can evolve for an arbitrary amount of time at rate~1, and is then ``frozen'' (rate~0).
}
Recently, a growing number of analyses and techniques were proposed to analyze parametric timed models (mainly PTAs) such as
	SMT-based techniques~\cite{KP12},
	integer hull abstractions~\cite{JLR15},
	corner-point abstractions~\cite{BBLS15},
	distributed verification~\cite{ACN15},
% 	multicore model-checking~\cite{TODO:tcheques},
	NDFS-based synthesis~\cite{NPvdP18},
	machine learning~\cite{ALin17,LSGA17},
	etc.
However, despite some case studies informally shared between these works, there is a lack of a common basis to compare new tools and techniques in a fair manner.
Without a stable list of benchmarks publicly available, it is difficult to assess the efficiency of a new algorithm.

\paragraph{Contribution}
We present here a library of benchmarks containing academic and industrial case studies collected in the past few years from academic papers and industrial collaborations.
In addition, a focus is made on (possibly toy) examples known to be unsolvable using current state-of-the-art techniques, with the hope to encourage the development of new techniques to solve them.
Benchmarks are available online in the \imitator{} input format, and distributed using the GNU General Public License.
\reviewer{3}{``p2, The first time IMITATOR is introduced, it would be nice to say
more about the project and its history.''}

\paragraph{Related libraries}
The library most related to ours is that by Chen \etal{}, that proposes a suite of benchmarks for hybrid systems~\cite{CSBMAFK15}.
However, it aims at analyzing hybrid systems, which are strictly more expressive than PTAs in theory, and incomparable in practice, as most hybrid systems do not feature timing parameters.
In addition, that benchmark suite focuses only on reachability properties.
Finally and most importantly, it does not focus on parameters, and the benchmarks are non-parametric.
In contrast, our library focuses on parametric timed benchmarks, with various types of properties.

Another interesting library is that by Hoxha, Abbas, and Fainekos~\cite{HAF14}, that offers Matlab/Simulink models of automotive systems.
However, it does not aim specifically at parametric timed model checking; two of our benchmarks originally partially come from the aforementioned library~\cite{HAF14}.

%%%%%%%%%%%%%%%%%%%%%%%%%%%%%%%%%%%%%%%%%%%%%%%%%%%%%%%%%%%%
%%%%%%%%%%%%%%%%%%%%%%%%%%%%%%%%%%%%%%%%%%%%%%%%%%%%%%%%%%%%
\section{\imitator{} parametric timed automata}\label{section:PTAs}
%%%%%%%%%%%%%%%%%%%%%%%%%%%%%%%%%%%%%%%%%%%%%%%%%%%%%%%%%%%%
%%%%%%%%%%%%%%%%%%%%%%%%%%%%%%%%%%%%%%%%%%%%%%%%%%%%%%%%%%%%

Parametric timed automata extend finite-state automata with clocks, \ie{} real-valued variables evolving at the same rate.
Clocks can be reset along transitions, and can be compared to constants or parameters (integer- or rational-valued) along transitions (``guards'') or in locations (``invariants'').
\imitator{} parametric timed automata extend PTAs~\cite{AHV93} with some useful features such as synchronization between components, stopwatches (\ie{} the ability to stop the elapsing of some clocks~\cite{CL00}), presence of parametric linear terms in guards, invariants and resets, shared global rational-valued variables, etc.

%%%%%%%%%%%%%%%%%%%%%%%%%%%%%%%%%%%%%%%%%%%%%%%%%%%%%%%%%%%%
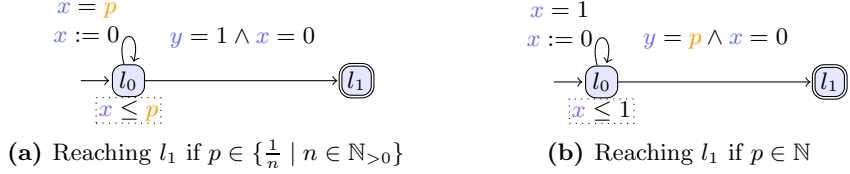
\begin{figure*}[t]
	\begin{subfigure}[b]{.49\textwidth}
	\centering
		\scalebox{1}{
		\begin{tikzpicture}[shorten >=1pt,node distance=3cm,on grid,auto]
		%% states
		\node[location,initial] (l0) {$\loc_0$};
		\node[location,accepting] (l1) [right of=l0] {$\loc_1$};
% 		\node[invariant] at l_0.south {plop};
		\node [invariant,below] at (l0.south) {$\styleclock{x} \leq \styleparam{p}$};

		%% edges
		\path[->] 
		(l0) edge [above] node[align=center]{
									$\styleclock{y} = 1 \land \styleclock{x} = 0$\\
% 									$\styleact{a}$\\
% 									$\styleclock{x} := 0$
								} (l1)
		(l0) edge [loop above] node[left, align=center]{
			$\styleclock{x} = \styleparam{p}$\\
			$\styleclock{x} := 0$\\
		} (l0)
% 		(s_3) edge [above] node[align=center] {$\CTrue$\\$\styleact{\$}$} (s_4)
		;
		\end{tikzpicture}}
	\caption{Reaching $\loc_1$ if $\param \in \{ \frac{1}{n} \mid n \in \grandn_{> 0} \}$\SpringerVersion{.}}
	\label{figure:example:PTA:1surn}
	\end{subfigure}
	\hfill{}
	\begin{subfigure}[b]{.49\textwidth}
	\centering
		\scalebox{1}{
		\begin{tikzpicture}[shorten >=1pt,node distance=3cm,on grid,auto]
		%% states
		\node[location,initial] (l0) {$\loc_0$};
		\node[location,accepting] (l1) [right of=l0] {$\loc_1$};
% 		\node[invariant] at l_0.south {plop};
		\node [invariant,below] at (l0.south) {$\styleclock{x} \leq 1$};

		%% edges
		\path[->] 
		(l0) edge [above] node[align=center]{
									$\styleclock{y} = \styleparam{\param} \land \styleclock{x} = 0$\\
% 									$\styleact{a}$\\
% 									$\styleclock{x} := 0$
								} (l1)
		(l0) edge [loop above] node[left, align=center]{
			$\styleclock{x} = 1$\\
			$\styleclock{x} := 0$\\
		} (l0)
% 		(s_3) edge [above] node[align=center] {$\CTrue$\\$\styleact{\$}$} (s_4)
		;
		\end{tikzpicture}}
	\caption{Reaching $\loc_1$ if $\param \in \grandn$\SpringerVersion{.}}
	\label{figure:example:PTA:n}
	\end{subfigure}

	\caption{Examples of PTAs\SpringerVersion{.}}
	\label{figure:example}
\end{figure*}
%%%%%%%%%%%%%%%%%%%%%%%%%%%%%%%%%%%%%%%%%%%%%%%%%%%%%%%%%%%%

\begin{example}
	Consider the PTA in \cref{figure:example:PTA:1surn}, containing two locations $\loc_0$ and~$\loc_1$, two clocks \styleclock{x} and \styleclock{y}, and one parameter~\styleparam{p}.
	The self-loop on~$\loc_0$ can be taken whenever $\styleclock{x} = \styleparam{p}$ holds, and resets $\styleclock{x}$, \ie{} can be taken every \styleparam{p} time units.
	In addition, initially, as $\styleclock{x} = \styleclock{y} = 0$ and clocks evolve at the same rate, the transition guarded by $\styleclock{y} = 1 \land \styleclock{x} = 0$ cannot be taken.
	Observe that, if $\styleparam{p} = 1$, then the transition to~$\loc_1$ can be taken after exactly one loop on~$\loc_0$.
	If $\styleparam{p} = \frac{1}{2}$, then the transition to~$\loc_1$ can be taken after exactly two loops.
	In fact, the set of valuations for which $\loc_1$ is reachable is exactly $\{ i \mid i = \frac{1}{n}, n > 0 \land n \in \grandn \}$.
\end{example}

\paragraph{L/U-PTAs}
Lower-bound/upper-bound parametric timed automata (L/U-PTAs) \cite{HRSV02} restrict the use of parameters: parameters must be partitioned between lower-bound parameters (always compared with clocks as lower bounds, \ie{} $\param \leq \clock$ or $\param < \clock$) and upper-bound parameters.
L/U-PTAs enjoy monotonicity properties and, while the full class of PTAs is highly undecidable~\cite{Andre18STTT}, L/U-PTAs enjoy some decidability results~\cite{HRSV02,BlT09,ALR18ACSD}.
U-PTAs~\cite{BlT09,ALR18FORMATS} are L/U-PTAs with only upper-bound parameters.

%%%%%%%%%%%%%%%%%%%%%%%%%%%%%%%%%%%%%%%%%%%%%%%%%%%%%%%%%%%%
%%%%%%%%%%%%%%%%%%%%%%%%%%%%%%%%%%%%%%%%%%%%%%%%%%%%%%%%%%%%
\section{The benchmark library}\label{section:benchmarks}
%%%%%%%%%%%%%%%%%%%%%%%%%%%%%%%%%%%%%%%%%%%%%%%%%%%%%%%%%%%%
%%%%%%%%%%%%%%%%%%%%%%%%%%%%%%%%%%%%%%%%%%%%%%%%%%%%%%%%%%%%

%%%%%%%%%%%%%%%%%%%%%%%%%%%%%%%%%%%%%%%%%%%%%%%%%%%%%%%%%%%%
\subsection{Categories}
%%%%%%%%%%%%%%%%%%%%%%%%%%%%%%%%%%%%%%%%%%%%%%%%%%%%%%%%%%%%

Our benchmarks are classified into three main categories:
\begin{enumerate}
	\item academic benchmarks, studied in a range of papers: a typical example is the Fischer mutual exclusion protocol;
	\item industrial case studies, which correspond to a concrete problem solved (or not) in an industrial environment;
	\item examples famous for being unsolvable using state-of-the-art techniques; for some of them, a solution may be computed by hand, but existing automated techniques are not capable of computing it.
		This is the case of the PTA in \cref{figure:example:PTA:1surn}, as a human can very easily solve it, while (to the best of our knowledge) no tool is able to compute this result automatically.
\end{enumerate}

\begin{remark}
	Our library contains a fourth category: education benchmarks, that consist of generally simple case studies that can be used for \emph{teaching}.
	This category contains toy examples such as coffee machines.
	We omit this category from this paper as these benchmarks generally have a limited interest performance wise.
\end{remark}

The domain of the benchmarks are hardware asynchronous circuits, communication or mutual exclusion protocols, real-time systems (``RTS'') and schedulability problems, parametric timed pattern matching (``PTPM''), train-gate-controllers models (``TGC''), etc.

In addition, we use the following classification criteria:
\begin{itemize}
	\item number of variables: clocks, parameters, locations, automata;
	\item whether the benchmark (in the provided version) is easily \emph{scalable}, \ie{} whether one can generate a large number of instances; for example, protocols often depend on the number of participants, and can therefore be scaled accordingly;
	\item presence of shared rational-valued variables;
	\item presence of stopwatches;
	\item presence of location invariants, as some works (\eg{} \cite{AHV93,ALR18FORMATS}) exclude them;
	\item whether the benchmark meets the L/U assumption.
% 	\item whether the symbolic zone graph\todo{} is finite;
\end{itemize}

%%%%%%%%%%%%%%%%%%%%%%%%%%%%%%%%%%%%%%%%%%%%%%%%%%%%%%%%%%%%
\subsection{Properties}
%%%%%%%%%%%%%%%%%%%%%%%%%%%%%%%%%%%%%%%%%%%%%%%%%%%%%%%%%%%%

We consider the three following main properties:
\begin{description}
	\item[reachability / safety:] synthesize parameter valuations for which a given state of the system (generally a location, but possibly a constraint on variables) must be reachable / avoided (see \eg{} \cite{JLR15}).
	\item[optimal reachability:] same as reachability, but with an optimization criterion: some parameters (or the time) should be minimized or maximized.
	\item[unavoidability:] synthesize parameter valuations for which all runs must always eventually reach a given state (see \eg{} \cite{JLR15}).
	\item[robustness:] synthesize parameter valuations preserving the discrete behavior (untimed language) \wrt{} to a given valuation (see \eg{} \cite{ACEF09,Sankur15}).
\end{description}
In addition, we include some recent case studies of parametric timed pattern matching (``PTPM'' hereafter), \ie{} being able to decide for which part of a log and for which values of parameters does a parametric property holds on that log~\cite{AHW18}.
Finally, a few more case studies have \emph{ad-hoc} properties (liveness, properties expressed using observers~\cite{ABBL98,Andre13ICECCS}, etc.), denoted ``Misc.'' later on.

%%%%%%%%%%%%%%%%%%%%%%%%%%%%%%%%%%%%%%%%%%%%%%%%%%%%%%%%%%%%
\subsection{Presentation}
%%%%%%%%%%%%%%%%%%%%%%%%%%%%%%%%%%%%%%%%%%%%%%%%%%%%%%%%%%%%

The benchmark library comes in the form of a Web page that classifies models %that can be found directly in the \imitator{} repository on Github\footnote{\url{https://github.com/imitator-model-checker/imitator/tree/master/benchmarks}}.
% The library
and
is available at \url{https://www.imitator.fr/library.html}.

The library is made of a list of a set of \emph{benchmarks}.
Each benchmark may have different \emph{models}: for example, \styleBen{Flip-flop} comes with three models, one with 2~parameters, one with 5, and one with 12~parameters.
Similarly, some \styleBen{Fischer} benchmarks come with several models, each of them corresponding to a different number of processes.
Finally, each model comes with one or more \emph{properties}.
For example, for \styleBen{Fischer}, one can either run safety synthesis, or evaluate the robustness of a given reference parameter valuation.

The first version of the library contains 34 benchmarks with 80 different models and 122 properties.

%%%%%%%%%%%%%%%%%%%%%%%%%%%%%%%%%%%%%%%%%%%%%%%%%%%%%%%%%%%%
\subsection{Performance}
%%%%%%%%%%%%%%%%%%%%%%%%%%%%%%%%%%%%%%%%%%%%%%%%%%%%%%%%%%%%

We present a selection of the library in \cref{table:library}.
Not all benchmarks are given; in addition, most benchmarks come with several models and several properties, omitted here for space concern.
% For example, for \styleBen{Jobshop}, \imitator{} is able to solve the case study for a smaller model (omitted here but present in the library).
We give from left to right the number of automata, of clocks, of parameters, of discrete variables, whether the model is an L/U-PTA, a U-PTA or a regular PTA, whether it features invariants and stopwatches, the kind of property, and a computation time on an Intel i7-7500U CPU @ 2.70GHz with 8\,GiB running Linux Mint~18.

\newcommand{\cellLU}{\cellcolor{green!20}L/U}
\newcommand{\cellU}{\cellcolor{green!20}U}
\newcommand{\cellNotLU}{\cellcolor{red!20}\textbf{$-$}}

\newcommand{\cellUnsolvable}{\cellcolor{red!20}?}
\newcommand{\cellHSolvable}{\cellcolor{red!20}HS}
\newcommand{\cellTO}{\cellcolor{red!20}T.O.}

\newcommand{\columnFinite}[1]{}

\begin{table}[tb!]
% 	\centering
	\caption{A selection from the benchmark library} % the parametric timed model checking 
	\noindent
	\scalebox{.87}{
	\begin{tabular}{@{}| c | c | c | c | c | c | c | c | c | c | c | c | r | }
		\hline
		\rowHeader{}
		Benchmark & Ref & Domain & Scal. & $|\A|$ & $|\Clock|$ & $|\Param|$ & $|\Variable|$ & L/U & Inv & SW\columnFinite{ & Finite} & Prop. & Time \\
		\hline
		%------------------------------------------------------------
		% BEGIN ACADEMIC
		\multicolumn{13}{| c |}{\cellSubHeader{}Academic}\\
		\hline
		\styleBen{And-Or} & \cite{CC05} & Circuit & \cellNo{} & 4 & 4 & 12 & 0 & \cellNotLU{} & \cellYes{} & \cellNo{} \columnFinite{ & \cellNo{}} & Misc. & 3.01 \\ % TODO: \todo{pattern / cycle}
		\hline
		\styleBen{CSMA/CD} & \cite{KNSW07} & Protocol & \cellYes{} & 3 & 3 & 3 & 0 & \cellNotLU{} & \cellYes{} & \cellNo{} \columnFinite{ & \cellNo{}} & Unavoid. & \cellUnsolvable{} \\ % TODO: + Reachability / 
		\hline
		\styleBen{Fischer-AHV93} & \cite{AHV93} & Protocol & \cellYes{} & 3 & 2 & 4 & 0 & \cellLU{} & \cellNo{} & \cellNo{} \columnFinite{ & \cellNo{}} & Safety & 0.04 \\
		\hline
% 		\styleBen{Fischer-HRSV02:2} & \cite{HRSV02} & Protocol & \cellYes{} & 2 & 2 & 4 & 1 & \cellLU{} & \cellYes{} & \cellNo{} \columnFinite{ & \cellNo{}} & Safety & \cellHSolvable{}\\
% 		\hline
		\styleBen{Fischer-HRSV02:3} & \cite{HRSV02} & Protocol & \cellYes{} & 3 & 3 & 4 & 1 & \cellLU{} & \cellYes{} & \cellNo{} \columnFinite{ & \cellNo{}} & Safety & \cellHSolvable{} \\
		\hline
		\styleBen{Flip-flop:2} & \cite{CC07} & Circuit & \cellNo{} & 5 & 5 & 2 & 0 & \cellU{} & \cellYes{} & \cellNo{} \columnFinite{ & \todo{Finite}} & Misc. & 0.04 \\
		\hline
		\styleBen{Flip-flop:12} & \cite{CC07} & Circuit & \cellNo{} & 5 & 5 & 12 & 0 & \cellU{} & \cellYes{} & \cellNo{} \columnFinite{ & \todo{Finite}} & Misc. & 23.07 \\
		\hline
		\styleBen{idle-time-sched:3} & \cite{LSAF14} & RTS & \cellYes{} & 8 & 13 & 2 & 3 & \cellU{} & \cellYes{} & \cellYes{} \columnFinite{ & \cellNo{}} & Safety & 1.49 \\
		\hline
		\styleBen{idle-time-sched:5} & \cite{LSAF14} & RTS & \cellYes{} & 12 & 21 & 2 & 0 & \cellU{} & \cellYes{} & \cellYes{} \columnFinite{ & \cellNo{}} & Safety & 14.61 \\
		\hline
		\styleBen{Jobshop:3-4} & \cite{AM01} & Sched. & \cellYes{} & 2 & 3 & 12 & 4 & \cellNotLU{} & \cellYes{} & \cellNo{} \columnFinite{ & \cellYes{}} & Opt. reach. & 5.58 \\
		\hline
		\styleBen{Jobshop:4-4} & \cite{AM01} & Sched. & \cellYes{} & 4 & 4 & 16 & 4 & \cellNotLU{} & \cellYes{} & \cellNo{} \columnFinite{ & \cellYes{}} & Opt. reach. & \cellTO{} \\
		\hline
		\styleBen{NP-FPS-3tasks:50-0} & \cite{JLR13} & RTS & \cellNo{} & 4 & 6 & 2 & 0 & \cellNotLU{} & \cellYes{} & \cellNo{} \columnFinite{ & \cellNo{}} & Safety & 1.03 \\
		\hline
		\styleBen{NP-FPS-3tasks:100-2} & \cite{JLR13} & RTS & \cellNo{} & 4 & 6 & 2 & 0 & \cellNotLU{} & \cellYes{} & \cellNo{} \columnFinite{ & \cellNo{}} & Safety & 65.23 \\
		\hline
% 		\styleBen{Packaging} & \cite{toDO} & Prod.-cons. & \cellNo{} & 3 & 2 & 2 & 0 & \cellLU{} & \cellYes{} & \cellNo{} \columnFinite{& \cellNo{}} & Reach. & \todo{Time} \\
% 		\hline
		\styleBen{SSLAF14-1} & \cite{PG98,SSLAF13} & RTS & \cellNo{} & 7 & 16 & 2 & 2 & \cellNotLU{} & \cellYes{} & \cellYes{} \columnFinite{ & \cellNo{}} & Safety & 0.33 \\
		\hline
		\styleBen{SSLAF14-2} & \cite{WTVL06,SSLAF13} & RTS & \cellNo{} & 6 & 14 & 2 & 4 & \cellNotLU{} & \cellYes{} & \cellYes{} \columnFinite{ & \cellNo{}} & Safety & \cellTO{} \\
		\hline
		\styleBen{ProdCons:2-3} & \cite{KP12} & Prod.-cons. & \cellYes{} & 5 & 5 & 6 & 0 & \cellLU{} & \cellYes{} & \cellNo{} \columnFinite{ & \cellNo{}} & Reach. & \cellTO{} \\
		\hline
		\styleBen{train-AHV93} & \cite{AHV93} & TGC & \cellNo{} & 3 & 3 & 6 & 0 & \cellLU{} & \cellNo{} & \cellNo{} \columnFinite{ & \cellNo{}} & Safety & 0.01 \\
		\hline
		\styleBen{WFAS} & \cite{BBLS15} & Protocol & \cellNo{} & 3 & 4 & 2 & 0 & \cellNotLU{} & \cellYes{} & \cellNo{} \columnFinite{ & \cellNo{}} & Safety & \cellTO{} \\
		\hline
		%------------------------------------------------------------
		% BEGIN INDUSTRIAL
		\multicolumn{13}{| c |}{\cellSubHeader{}Industrial}\\
		\hline
		\styleBen{accel:1} & \cite{HAF14,AHW18} & PTPM & \cellYes{} & 2 & 2 & 3 & 0 & \cellNotLU{} & \cellYes{} & \cellNo{} \columnFinite{ & \cellYes{}} & PTPM & 1.25 \\
		\hline
		\styleBen{accel:10} & \cite{HAF14,AHW18} & PTPM & \cellYes{} & 2 & 2 & 3 & 0 & \cellNotLU{} & \cellYes{} & \cellNo{} \columnFinite{ & \cellYes{}} & PTPM & 12.67 \\
		\hline
		\styleBen{BRP} & \cite{DKRT97} & Protocol & \cellNo{} & 6 & 7 & 2 & 12 & \cellNotLU{} & \cellYes{} & \cellNo{} \columnFinite{ & \cellNo{}} & Safety & 248.35 \\
		\hline
		\styleBen{FMTV:1A1} & \cite{SAL15} & RTS & \cellNo{} & 3 & 3 & 3 & 5 & \cellNotLU{} & \cellYes{} & \cellNo{} \columnFinite{ & \cellNo{}} & Opt. reach. & 6.97 \\
		\hline
		\styleBen{FMTV:1A3} & \cite{SAL15} & RTS & \cellNo{} & 3 & 3 & 3 & 7 & \cellNotLU{} & \cellYes{} & \cellNo{} \columnFinite{ & \cellNo{}} & Opt. reach. & 87.39 \\
		\hline
		\styleBen{FMTV:2} & \cite{SAL15} & RTS & \cellNo{} & 6 & 9 & 2 & 0 & \cellNotLU{} & \cellYes{} & \cellYes{} \columnFinite{ & \cellNo{}} & Opt. reach. & 1.61 \\
		\hline
		\styleBen{gear:1} & \cite{HAF14,AHW18} & PTPM & \cellYes{} & 2 & 2 & 3 & 0 & \cellNotLU{} & \cellYes{} & \cellNo{} \columnFinite{ & \cellYes{}} & PTPM & 0.77 \\
		\hline
		\styleBen{gear:10} & \cite{HAF14,AHW18} & PTPM & \cellYes{} & 2 & 2 & 3 & 0 & \cellNotLU{} & \cellYes{} & \cellNo{} \columnFinite{ & \cellYes{}} & PTPM & 7.42 \\
		\hline
		\styleBen{RCP} & \cite{CS01} & Protocol & \cellNo{} & 5 & 6 & 5 & 6 & \cellLU{} & \cellYes{} & \cellNo{} \columnFinite{ & \cellNo{}} & Reach. & 1.07 \\
		\hline
		\styleBen{SIMOP:3} & \cite{ACDFR09} & Automation & \cellNo{} & 5 & 8 & 3 & 0 & \cellNotLU{} & \cellYes{} & \cellNo{} \columnFinite{ & \cellNo{}} & Reach. & \cellTO{} \\
		\hline
		\styleBen{SPSMALL:2} & \cite{CEFX09} & Circuit & \cellNo{} & 11 & 11 & 2 & 0 & \cellNotLU{} & \cellYes{} & \cellNo{} \columnFinite{ & ???} & Reach. & 0.96 \\
		\hline
		\styleBen{SPSMALL:26} & \cite{CEFX09} & Circuit & \cellNo{} & 11 & 11 & 26 & 0 & \cellNotLU{} & \cellYes{} & \cellNo{} \columnFinite{ & ???} & Reach. & \cellTO{} \\
		\hline
		%------------------------------------------------------------
		% BEGIN TOY
		\multicolumn{13}{| c |}{\cellSubHeader{}Toy}\\
		\hline
		\styleBen{toy:n} & \cref{figure:example:PTA:n} & Toy & \cellNo{} & 1 & 2 & 1 & 0 & \cellNotLU{} & \cellYes{} & \cellNo{} \columnFinite{& \cellNo{}} & Reach. & \cellHSolvable{} \\
		\hline
		\styleBen{toy:1/n} & \cref{figure:example:PTA:1surn} & Toy & \cellNo{} & 1 & 2 & 1 & 0 & \cellU{} & \cellYes{} & \cellNo{} \columnFinite{& \cellNo{}} & Reach. & \cellHSolvable{} \\
		\hline
% 		\styleBen{Benchmark} & \cite{toDO} & \todo{Domain} & \todo{Scalable} & \todo{$|\A|$} & \todo{$|\Clock|$} & \todo{$|\Param|$} & \todo{$|\Variable|$} & \todo{L/U} & \todo{Inv} & \todo{SW} \columnFinite{& \todo{Finite}} & \todo{Prop.} & \todo{Time} \\
% 		\hline
		
	\end{tabular}
	} % END scalebox
	
	\label{table:library}
\end{table}

% NOTE: j'ai supprimé TOUS les education + :
% - IMPO et variantes

\todo{Jobshop: et Romain!}

\todo{pattern matching = reachability}

``\cellTO{}'' denotes time-out (after 300\,s).
``\cellUnsolvable{}'' denotes unsolvable, because no such algorithm is implemented in existing tools.
``\cellHSolvable{}'' denotes time-out but human-solvable: \eg{} for Fischer, one knows the correctness constraint independently of the number of processes, but tools may fail to compute it.
This is also the case of the toy PTAs in \cref{figure:example:PTA:1surn,figure:example:PTA:n}.

Despite time-out, some case studies come with a partial result: either because \imitator{} is running reachability-synthesis (``\EFsynth{}''~\cite{JLR15}) which can output a partial result when interrupted before completion, or because some other methods can output some valuations.
For example, for \styleBen{ProdCons}, \imitator{} is unable to synthesize a constraint; however, in the original work~\cite{KP12}, some punctual valuations (non-symbolic) are given.

Robustness case studies are not part of \cref{table:library}, but are included in the online library.

%%%%%%%%%%%%%%%%%%%%%%%%%%%%%%%%%%%%%%%%%%%%%%%%%%%%%%%%%%%%
%%%%%%%%%%%%%%%%%%%%%%%%%%%%%%%%%%%%%%%%%%%%%%%%%%%%%%%%%%%%
\section{Perspectives}\label{section:conclusion}
%%%%%%%%%%%%%%%%%%%%%%%%%%%%%%%%%%%%%%%%%%%%%%%%%%%%%%%%%%%%
%%%%%%%%%%%%%%%%%%%%%%%%%%%%%%%%%%%%%%%%%%%%%%%%%%%%%%%%%%%%

\paragraph{Syntax}
So far, all benchmarks use the \imitator{} input format; in addition, only if the benchmark comes from another model checker (\eg{} a \hytech{} or \uppaal{} model), it also comes with its native syntax.
In a near future, we plan to propose a translation to \uppaal{} timed automata; however, some information will be lost as \uppaal{} does not allow parameters, and supports stopwatches in a limited manner.
A future work will be to propose other syntaxes, or a normalized syntax for parametric timed model checking benchmarks.

\paragraph{Contributions and versioning}
The library is aimed at being enriched with future benchmarks.
Furthermore, it is collaborative, and is open to any willing contributor.
A versioning system will be set up with the addition (or modification) of benchmarks in the future.

% 
% %%%%%%%%%%%%%%%%%%%%%%%%%%%%%%%%%%%%%%%%%%%%%%%%%%%%%%%%%%%%
% %%%%%%%%%%%%%%%%%%%%%%%%%%%%%%%%%%%%%%%%%%%%%%%%%%%%%%%%%%%%
% \section*{Acknowledgements}
% %%%%%%%%%%%%%%%%%%%%%%%%%%%%%%%%%%%%%%%%%%%%%%%%%%%%%%%%%%%%
% %%%%%%%%%%%%%%%%%%%%%%%%%%%%%%%%%%%%%%%%%%%%%%%%%%%%%%%%%%%%
% XXXXX

%%%%%%%%%%%%%%%%%%%%%%%%%%%%%%%%%%%%%%%%%%%%%%%%%%%%%%%%%%%%%
%%%%%%%%%%%%%%%%%%%%%%%%%%%%%%%%%%%%%%%%%%%%%%%%%%%%%%%%%%%%%
\ifdefined\VersionAuthor
	\bibliographystyle{alpha} % plain
	\newcommand{\IJFCS}{International Journal of Foundations of Computer Science}
	\newcommand{\JLAP}{Journal of Logic and Algebraic Programming}
	\newcommand{\LNCS}{Lecture Notes in Computer Science}
	\newcommand{\STTT}{International Journal on Software Tools for Technology Transfer}
	\newcommand{\ToPNoC}{Transactions on Petri Nets and Other Models of Concurrency}
\else
	\bibliographystyle{splncs04} % abbrv
	\newcommand{\IJFCS}{International Journal of Foundations of Computer Science}
	\newcommand{\JLAP}{Journal of Logic and Algebraic Programming}
	\newcommand{\LNCS}{Lecture Notes in Computer Science}
	\newcommand{\STTT}{International Journal on Software Tools for Technology Transfer}
	\newcommand{\ToPNoC}{Transactions on Petri Nets and Other Models of Concurrency}
\fi
\bibliography{benchmarks}
%%%%%%%%%%%%%%%%%%%%%%%%%%%%%%%%%%%%%%%%%%%%%%%%%%%%%%%%%%%%%
%%%%%%%%%%%%%%%%%%%%%%%%%%%%%%%%%%%%%%%%%%%%%%%%%%%%%%%%%%%%%

\end{document}